# Emergent anisotropy in the Fulde–Ferrell–Larkin–Ovchinnikov state

Shusaku Imajo [*], Toshihiro Nomura, Yoshimitsu Kohama, and Koichi Kindo
*Institute for Solid State Physics, University of Tokyo, Kashiwa 277-8581, Japan*
*E-mail: imajo@issp.u-tokyo.ac.jp

**Exotic superconductivity is formed by unconventional electron pairing and exhibits various unique properties that cannot be explained by the basic theory. The Fulde–Ferrell–Larkin–Ovchinnikov (FFLO) state is known as an exotic superconducting state in that the electron pairs have a finite center-of-mass momentum leading to a spatially modulated pattern of superconductivity. The spatial modulation endows the FFLO state with emergent anisotropy. However, the anisotropy has never been experimentally verified despite numerous efforts over the years. Here, we report detection of anisotropic acoustic responses depending on the sound propagation direction appearing above the Pauli limit. This anisotropy reveals that the two-dimensional FFLO state has a center-of-mass momentum parallel to the nesting vector on the Fermi surface. The present findings will facilitate our understanding of not only superconductivity in solids but also exotic pairings of various particles.**

## Introduction

As theorized by Bardeen, Cooper, and Schrieffer (BCS), superconductivity occurs when itinerant electrons form pairs, so-called Cooper pairs, via an attractive force. Although many superconducting properties are well described by BCS theory, long-standing studies have found various superconductors beyond the BCS framework and many intriguing open questions. One of the exotic unconventional superconducting states, the Fulde–Ferrell–Larkin–Ovchinnikov (FFLO) state, was independently theorized by Fulde & Ferrell (1) and Larkin & Ovchinnikov (2) in 1964. An up-spin electron with momentum $\mathbf{k}$ is coupled with a down-spin electron with momentum $-\mathbf{k}+\mathbf{q}$ in an FFLO pairing, leading to a finite center-of-mass momentum of Cooper pairs $\mathbf{q} \neq 0$, whereas an ordinary superconducting state is formed by electrons whose momenta are $\mathbf{k}$ and $-\mathbf{k}$, as illustrated in Fig. 1a. For ordinary superconductivity, in which spins in paired electrons are antiparallel to each other, a magnetic field destabilizes the superconductivity through the orbital effect and the Zeeman effect. In most superconductors, superconductivity is suppressed by the orbital effect caused by the Lorentz force on vortices, which mainly determines the upper critical field $H_{c2}$. However, when the orbital effect is quenched, the Zeeman effect governs $H_{c2}$. In this case, the ordinary spin-singlet superconducting state is destroyed at the field where the Zeeman splitting energy reaches the superconducting energy gap $\Delta$, known as the Pauli paramagnetic limit $H_P$. In contrast, the FFLO state is stable even above $H_P$ due to a gain in spin polarization energy of the nonzero $\mathbf{q}$. This finite $\mathbf{q}$ adds a term, $\cos(\mathbf{q}\mathbf{r})$, to the order parameter of the superconductivity $\Delta$. The modified gap function $\Delta\cos(\mathbf{q}\mathbf{r})$ indicates that the order parameter spatially oscillates in real space, as shown in Fig. 1b (3-8). The oscillatory pattern composed of the normal state and the superconductivity endows the FFLO state with emergent anisotropy depending on the $\mathbf{q}$ vector. Since disorder stunts the formation of the spatial modulation, the FFLO state appears only in the clean limit (4). Besides, as mentioned-above, the emergence of the FFLO state is allowed when the orbital effect is sufficiently weaker than the paramagnetic effect, as characterized by the Maki parameter $\alpha_M > 1.8$ (5). Consequently, these restrictions narrow the candidate materials in the search for FFLO superconductivity (7-14) and have disturbed experimental examination of FFLO physics despite numerous theoretical studies (1-6,15-19). In particular, spatial anisotropy, one of the main features of the FFLO state, has never been experimentally observed.

The organic superconductor $\kappa$-(BEDT-TTF)$_2$Cu(NCS)$_2$ (BEDT-TTF is an abbreviation of bis(ethyleneditio)tetrathiafulvalene) is known as the prime candidate for exhibiting the FFLO state and has been examined by various measurements (9,20-26). As displayed in Fig. 1c, the layered structure formed by alternating stacking of conducting and insulating layers provides a quasi-two-dimensional (quasi-2D) electronic structure. This compound undergoes a superconducting transition at ~9.5 K and changes into a $d$-wave superconductivity (27,28), which is theoretically expected to manifest using nesting vectors including the predominant nesting vector $\mathbf{Q_1}$ shown in Fig. 1d (29,30).

When a magnetic field is parallel to the conducting plane, emergence of FFLO pairing is highly expected in κ-(BEDT-TTF)$_2$Cu(NCS)$_2$ because of the large Maki parameter, relatively long mean-free path (9), and quasi-2D Fermi surface. The heat capacity data (21) show a 1st-order transition at $H_{FFLO}$~21 T. Tunnel diode oscillator (TDO) measurement (22) and torque magnetometry (23,24) also detect this anomaly, which is smeared out by a slight field misalignment. Based on the existence of an additional superconducting phase with an upturn in its field-temperature phase diagram, these works suggest that a putative FFLO phase appears in the high-field region. Moreover, a NMR study (26) detects the formation of Andreev bound states related to a phase twist of the order parameter, which strongly indicates the presence of the FFLO phase (31). A slight in-plane anisotropy in $H_{c2}$ (23) implies that the FFLO state has a $d$-wave-like fourfold gap symmetry in momentum space. However, conclusive evidence for anisotropy related to spatial modulation in real space is still missing. Employing multidirectional ultrasound propagation, we examine the plausible FFLO state of κ-(BEDT-TTF)$_2$Cu(NCS)$_2$ and establish that this state certainly exhibits an anisotropic response, which is a hallmark of the FFLO state.

**Results**

To discuss the anisotropy in the FFLO state, we arranged two pairs of transducers, generating and detecting longitudinal ultrasonic waves, on all sides of a cuboid-shaped single crystal, as shown in Fig. 1e. First, in Fig. 1f, we show the relative change in sound velocity $\Delta v/v$ and ultrasonic attenuation $\Delta\alpha$ at 1.6 K in magnetic fields perpendicular to the conducting plane $\theta$=90°. The polarization vector **u** (parallel to the ultrasound propagation vector for longitudinal waves) is along the $b$-axis. At low fields, the obtained data reproduce the reported behavior (32). As indicated by the arrow, $\Delta v/v$ exhibits an anomaly accompanied by suppression of the superconductivity at 3 T (=$H_{c2}$(90°)). The lattice softening in most of its superconducting region in a perpendicular field (Fig. 1f) agrees with the fact that the vortex lattice melts at a much lower field (<0.5 T) when $H \| a^*$ (33). From the equation $H_{c2}$(90°)=$\varphi_0/(2\pi\xi_\|^2)$, where $\varphi_0$ is the flux quantum, an in-plane coherence length of $\xi_\|$~10 nm is determined. The gradual increase in $\Delta v/v$ between 3 T and 7 T reflects the suppression of fluctuating superconductivity above $H_{c2}$ (34) since $\Delta v/v$ is a sensitive indicator of fluctuations of the superconducting order parameter in organic superconductors (32,35). At higher fields, both properties exhibit the acoustic de Haas–van Alphen (acoustic dHvA) oscillations mainly composed of two orbits, whose frequencies are estimated to be approximately 610 T and 3300 T. The obtained frequencies well coincide with the reported values of the α orbit (blue area in Fig. 1d) and the β−α orbit (36,37). Detailed analyses and discussions are described in the Supplementary Materials. For the α orbit, an estimation of the mean-free path $l$ from a fit to the typical Lifshitz-Kosevich formula leads to $l$~90 nm. This value sufficiently larger than $\xi_\|$~10 nm indicates that the electronic system is in the clean limit, which meets one of the requirements for the emergence of the FFLO state. For the Maki parameter, the phase diagram, which is consistent with our results discussed later, indicates $\alpha_M$~8 (9), which is approximately 4 times larger than required (5). Thus, the present sample satisfies the conditions required to form the FFLO state.

Since the FFLO state appears at low temperatures when the orbital effect is sufficiently suppressed, in Fig. 2, we show the magnetic field dependence of the elastic properties ($\mathbf{u} \| b$) at 2.1 K near the parallel direction, with $\theta < 1.2°$. Note that no clear hysteresis was observed in our present measurements (see Supplementary Fig. 4). At $\theta = 0°$, the field dependence of $\Delta v/v$ has two dips at ~21.3 T (blue circle) and ~24.5 T (black triangle), as indicated by the symbols in Fig. 2a. These anomalies are observed as peaks in $\Delta \alpha$ in Fig. 2b. Based on the results of previous studies (9,20-26), these characteristic fields correspond to $H_{FFLO}$ and $H_{c2}$, respectively, and the FFLO state appears between $H_{FFLO}$ and $H_{c2}$. Upon tilting the sample away from 0°, $H_{c2}$ abruptly decreases, whereas $H_{FFLO}$ shows barely any change. Since these two anomalies finally merge into one sharp anomaly at $\theta = 1.2°$, the FFLO state is completely suppressed by this slight tilt. This result is perfectly consistent with the report that the FFLO state at 2.0 K only exists for $\theta < 1.2°$ (23). To closely examine the ultrasonic properties of the FFLO state, the datasets of $\theta = 0°$ and $\theta = 1.2°$ are enlarged in Fig. 2c and 2d. The light green area corresponds to the contribution of the FFLO state. This result indicates that the formation of the FFLO state leads to the lattice hardening in the $\mathbf{u} \| b$ direction. For $\Delta \alpha$ in Fig. 2d, attenuation of the sound wave propagation by FFLO formation is natural because of the spatially inhomogeneity. Near $H_{c2}$, a flux flow gives excess attenuation appearing as a peak in $\Delta \alpha$ (38,39), and therefore, the combination of the two peaks at $H_{FFLO}$ and $H_{c2}$ produces the observed behavior above $H_{FFLO}$. Note that the difference below $H_{FFLO}$ (gray area) originates from perpendicular components of the applied fields because it appears when a field is tilted away from $\theta = 0°$. The perpendicular component, which penetrates the conducting plane and forms pancake vortices, induces excess dynamics of the pancake vortices. Therefore, when $\theta \neq 0°$, the lattice is softened and $\Delta \alpha$ is enhanced, leading to the difference highlighted by the gray area.

In Fig. 3a and 3b, we show the $H_{FFLO}$ and $H_{c2}$ of the detected anomalies at 2.1 K and 6.3 K as an $H$ vs. $\theta$ plot. The cusp-like angular dependence of $H_{c2}$ at 6.3 K can be described by the Tinkham 2D model (40). Indeed, the interlayer coherence length $\xi_\perp \sim 1.4$ nm ($= \varphi_0 / 2\pi \xi_\| H_{c2}(0°)$) is smaller than the interlayer distance of 1.5 nm, indicating that the interlayer coupling of the superconductivity is weak. This anisotropic behavior and the values of $H_{c2}$ agree well with the results reported in Ref. (22,24,33). However, this model cannot reproduce $H_{c2}$ at 2.1 K because of the emergence of the FFLO state. The $H_{c2}$ determined by the resistivity (see Supplementary Materials) at 1.6 K also exhibits similar behavior, as shown by the pink triangles (right axis) in Fig. 3a. The abrupt suppression of $H_{c2}$ when moving away from $\theta = 0°$ means that the FFLO state is easily destabilized even by the small orbital effect induced by the slight tilt. This fragility to the orbital effect is also a well-known characteristic peculiar to the FFLO state (5,18,41). In contrast, the angle dependence of $H_{FFLO}$ is not significant. As $H_{FFLO}$ corresponds to $H_P$ determined by the paramagnetic effect, the angle-insensitive behavior is suggestive of isotropic Pauli paramagnetism. This fact is also consistent with the almost isotropic $g$-factor in the organic compounds composed of light atoms with weak spin-orbit coupling. In Fig. 3c, we organize the present results as the obtained $H$-$T$ superconducting phase diagram at $\theta$=0°. For comparison, we additionally show the data of earlier reports (blank symbols) (9,20-26). Our results are in good agreement with the reported data. In addition, the temperature dependence of the reduced superconducting gap amplitude $\Delta(T)/\Delta(0\ \text{K})$ calculated by the basic BCS theory is also shown on the right axis. Since the α model, a simple extension of the BCS theory, well describes the thermal variation in $|\Delta(T)/\Delta(0\ \text{K})|$ (42), the behavior roughly reconciling with the temperature dependence of $H_P$ for the homogeneous superconducting state is reasonable. Above we assumed that the FFLO state would appear between $H_{\text{FFLO}}$ and $H_{c2}$ according to the results of previous studies, this consistency certainly confirms that the high-field phase is non-BCS superconductivity emerging above $H_P$.

Apart from the phase diagram, examination in further detail of the pinning effect enhanced in the FFLO state is interesting. In Fig. 4a, we compare the field dependence of $\Delta v/v$ ($\theta$=0°) taken for the parallel ($\mathbf{u}\|c$) and perpendicular ($\mathbf{u}\|b$) configurations under the same conditions. There is only a small difference depending on the sound wave direction below $H_{\text{FFLO}}$. The difference becomes significantly larger in the FFLO region $H_{\text{FFLO}}$<$H$<$H_{c2}$ (green area). Since the acoustic response for $\theta$=90° is almost isotropic in the whole field region, as shown in Fig. 4b, the behavior is clear evidence of the emergent anisotropy of the FFLO state.

**Discussion**

Since the anisotropy appears only in the reported field-temperature region of the FFLO state, the present results demonstrate that the emergent anisotropy originates from the formation of the FFLO state. Given an additional periodicity of the spatial modulation, it seems natural that the stiffness of the lattice in the direction across that modulation pattern should increase. Thus, sound velocity measures the stiffness of crystal lattice. Namely, the $\mathbf{q}$ vector is oriented along the $b$-axis, perpendicular to the field direction in the present setup. Nevertheless, how this manifests as a change in $\Delta v/v$ must be discussed. The most likely possibility is flux pinning because vortices have a strong influence on elastic properties in the superconducting state (38,39,43). Typically, suppression of vortex motion results in lattice hardening through increase in spring constant. As shown in Fig. 2a, the enhancement of $\Delta v/v$ with increasing magnetic field in the lower-field region indicates compression of the flux-line lattice, which reduces the vortex dynamics. When the vortex lattice melts, $\Delta v/v$ decreases and shows a minimum. In Fig. 2c and 4a, the lattice hardening observed in $\mathbf{u}\|b$ suggests that the spatial modulation when $H\|c$ reinforces pinning of the flux lines. When this spatial modulation traps the flux lines, the Josephson vortices should be pinned at nodes of the spatial modulation of the FFLO state because the Josephson supercurrent is absent at the node positions. Therefore, for the strong pinning effect, the wavelength of the order parameter oscillation $2\pi/q$ ($q$=$|\mathbf{q}|$) should be comparable to $d_{\text{JV}}/n$, using simple integers $n$ and the Josephson vortex lattice constant $d_{\text{JV}}$ ($d_{\text{JV}}$=$\varphi_0/sH$, where $s$ is the layer spacing) (41,44,45). Using the $a^*$-axis length as the layer spacing, 1.5 nm, $d_{JV}$ at 21 T is estimated to be approximately 60 nm. This estimation means that pinning by the FFLO formation requires $q \sim n*10^8$ m$^{-1}$. In the FFLO state, the energy gain by the momentum **q** is larger than the Zeeman splitting energy of the up- and down-spin Fermi surfaces. This energy balance gives the relation $q \hbar v_F = g \mu_B H_{FFLO}$. Here, $v_F$, $g$, and $\mu_B$ are the Fermi velocity, the $g$-value, and the Bohr magneton, respectively. This rough approximation leads to $q = 6*10^7$ m$^{-1}$ ($2\pi/q \sim 100$ nm) at 21 T. This estimation yields $n \sim 0.6$ at 21 T, which indicates the less strong pinning effect at 21 T. With increasing the field up to $H_{c2}$, the size of $2\pi/q$ abruptly decreases down to $\pi \xi_\parallel \sim 30$ nm (45,46), and thus, $n$ should pass through 1 and reaches about 2 near $H_{c2}$. The change in the pinning effect with the variation in $n$ results in the observed peak structure in the field dependence of $\Delta v/v$ in the FFLO state. The commensurability effect on the Josephson vortices (41,44,45) can be confirmed by the anomalies in the field dependence of the out-of-plane resistivity (see Supplementary Materials). Thus, the lattice hardening in $u \parallel b$ in the FFLO state may originate from the strong pinning effect of the flux-line lattice. Nevertheless, we should take into account other effects leading to the lattice hardening because the coupling between sound waves and Josephson-vortex lattice has not been well studied yet. For example, the change in the electron-lattice coupling and/or the local density of states related to the Andreev bound states may also modify the elastic properties. The present results certainly demonstrate the fact that the FFLO state yields the emergent anisotropy with the **q** vector along the $b$-axis; however, the detailed mechanism should be clarified by further measurements in the future.

Next, we need to consider the question of why the orientation of the **q** vector is mainly along the $b$-axis, perpendicular to the field direction in the present configuration. Note, here, that the effect of the spin-orbit interaction on the FFLO state in the present salt is negligible because the organic salt is composed of light elements. In the case of ideal isotropic 3D superconductors, the **q** vector always points in the field direction (1,2). Since the **q** vector can be oriented in any direction in 3D, the anisotropy can be discussed in the Heisenberg-type model. According to this framework, in the present measurement with $H \parallel c$, the **q** vector should be parallel to the $c$-axis, not the $b$-axis. However, the present superconductivity is described by the 2D model (Fig. 3b). For 2D superconductors, the better nesting vectors on the Fermi surface make the FFLO state more stable, and the anisotropy of the Fermi surface often locks the direction of the **q** vector according to the predominant nesting vectors (6-8,45). Namely, the FFLO state in the anisotropic 2D superconductor is expected to show Ising-type behavior. Indeed, theoretical studies (29,30) suggest that the nesting vector $\mathbf{Q_1}$, parallel to the $k_b$-axis (green arrow in Fig. 1d), always strongly relates to the Cooper pairing in $\kappa$-type organic salts regardless of the emergent pairing symmetry. Thus, the **q** vector parallel to the $b$-axis in the FFLO state should be reasonable for the present 2D superconductor. This result suggests that the direction of the emergent anisotropy and the model describing it, such as Ising, XY, and Heisenberg, can be controlled by changing the shape of the Fermi surface and dimensionality. Future studies of the in-plane field angle dependence of the emergent anisotropy will allow for further detailed discussions of the relation between the **q** vector and nesting vector. Furthermore, similar measurements in other FFLO candidates would also be interesting, as different anisotropies using other nesting vectors should occur in other FFLO candidates.

The present multidirectional ultrasound measurements demonstrate the emergent anisotropy of the FFLO state induced by the spatial modulation of the order parameter. Since $\kappa$-(BEDT-TTF)$_2$Cu(NCS)$_2$ is a 2D clean superconductor, the FFLO state shows Ising anisotropy originating from the anisotropic Fermi surface. Further studies of other FFLO candidates with various features, such as 3D and slight dirtiness, will facilitate a deeper understanding of the FFLO state.

## Acknowledgements

We thank R. Kurihara (Tokyo University of Science) and S. Sugiura (Tohoku University) for giving us fruitful advice. S.I. is supported by Japan Society for the Promotion of Science KAKENHI Grant 20K14406 and 22H04466.

## Methods

**Sample preparation**: Single crystals of $\kappa$-(BEDT-TTF)$_2$Cu(NCS)$_2$ measured in this study were synthesized by typical electrochemical process and crystallized as black hexagonally-shaped blocks. The shape of the crystals used in the ultrasonic measurements was modified as described in the Supplementary Materials.

**Ultrasonic measurements**: Using the typical pulse-echo methods, the ultrasonic properties were measured. Longitudinal ultrasound waves, whose frequencies were in the range of 37-39 MHz, were generated and detected by LiNbO$_3$ piezoelectric transducers (90 μm thickness) attached on side surfaces of the crystals. Further details of the setup are presented in the Supplementary Materials.

## Author Contributions

S.I. conceived the study and wrote the paper. S.I. synthesized the single crystals measured in the study. S.I. and K.K. constructed and prepared the measurement apparatus. S.I. and T.N. performed the measurements in pulsed magnetic fields. T.N. and Y.K. supervised the writing of the paper. All authors discussed the results and commented on the manuscript.

**Figure 1 Schematics of the FFLO state, crystal structure, and experimental setup.**

**a** Schematics of ordinary spin-singlet pairing and FFLO pairing. $\mathbf{k}$ and $\sigma$ represent the momentum and spin of the electrons.

**b** Spatial modulation of order parameter $\Delta(\mathbf{r})$ in the FFLO state (dashed curve). The normal state (blue) appears at nodes of the superconducting order parameter $\Delta\cos(\mathbf{qr})$.

**c** Quasi-two-dimensional crystal structure of $\kappa$-(BEDT-TTF)$_2$Cu(NCS)$_2$.

**d** Fermi surface of $\kappa$-(BEDT-TTF)$_2$Cu(NCS)$_2$. The solid and dashed black lines are the first Brillouin zone and the extended Brillouin zone, respectively. The blue and red curves show the Fermi surfaces, and the green arrow indicates the most predominant nesting vector.

**e** Experimental setup for the multidirectional ultrasound measurements using longitudinal sound waves. The magnetic field was rotated from the $c$-axis to the $a^*$-axis. $\theta$ is the polar angle from the $c$-axis.

**f** Relative change in the sound velocity $\Delta v/v$ (red, left axis) and attenuation coefficient $\Delta\alpha$ (blue, right axis) at $T$=1.6 K and $\theta$=90° as a function of magnetic field. The black arrow indicates the upper critical field of the superconductivity $H_{c2}$.

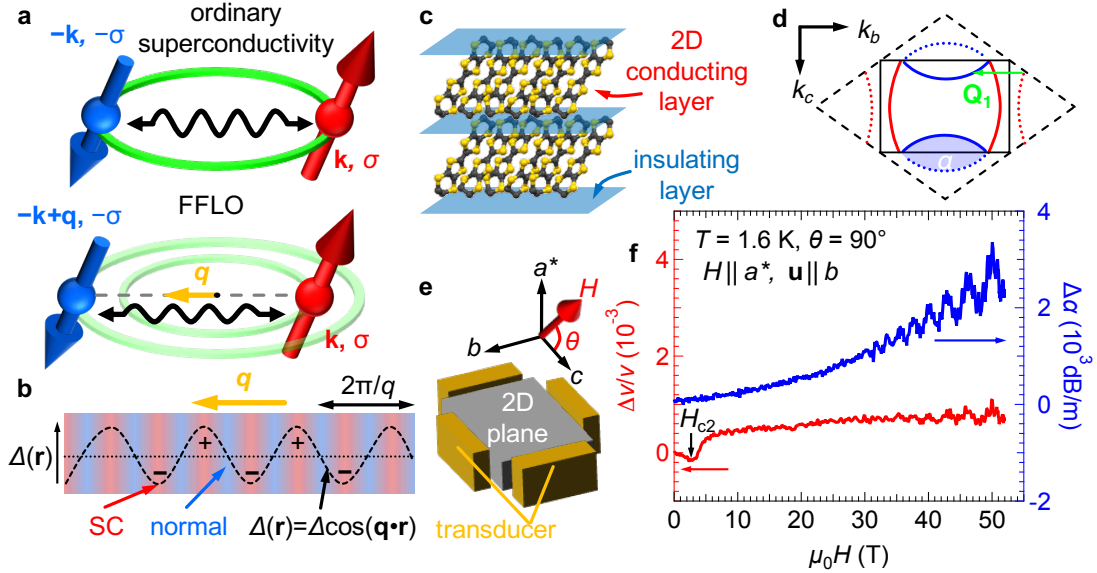

**Figure 2　Low-temperature and high-field ultrasonic properties when $\theta$–0°.**

**a,b** Magnetic field dependence of $\Delta v/v$ (**a**) and $\Delta\alpha$ (**b**) at 2.1 K. The blue and black symbols indicate the dips in $\Delta v/v$ and the peaks in $\Delta\alpha$. The black line represents offset for each dataset. **c,d** Enlarged plot of the datasets at 0° and 1.2°. The green area indicates the difference between the 0° and 1.2° data, which reflects the contribution of the FFLO state.

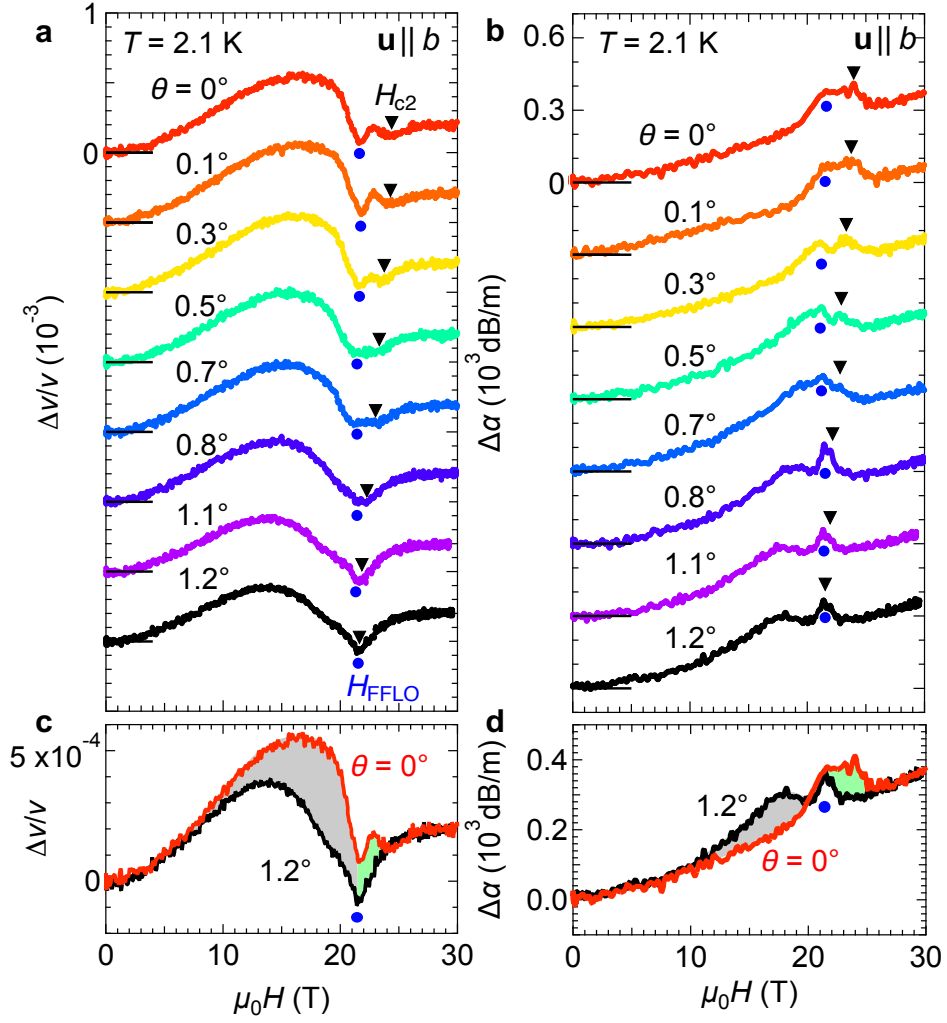

**Figure 3  Superconducting phase diagrams.**

**a,b** Upper critical field $H_{c2}$ (black triangles) as a function of the absolute value of the field angle $|\theta|$ at (**a**) 2.1 K and (**b**) 6.3 K. The blue circles in (**a**) indicate transition fields to the FFLO state $H_{FFLO}$. The green and red areas represent the regions for the FFLO phase and ordinary superconducting phase, respectively. The dotted curves in (**a**) and (**b**) are fits to the Tinkham 2D model, $|H_{c2}(\theta)\cos\theta/H_{c2}(90°)|+(H_{c2}(\theta)\sin\theta/H_{c2}(0°))^2=1$. **c** Field-temperature superconducting phase diagram obtained in this work (filled symbols) and previous reports (blank symbols) (9,21,22,26). The dashed curve (right axis) is the temperature dependence of the reduced superconducting gap amplitude $\Delta(T)/\Delta(0\ \text{K})$ calculated by the basic BCS theory.

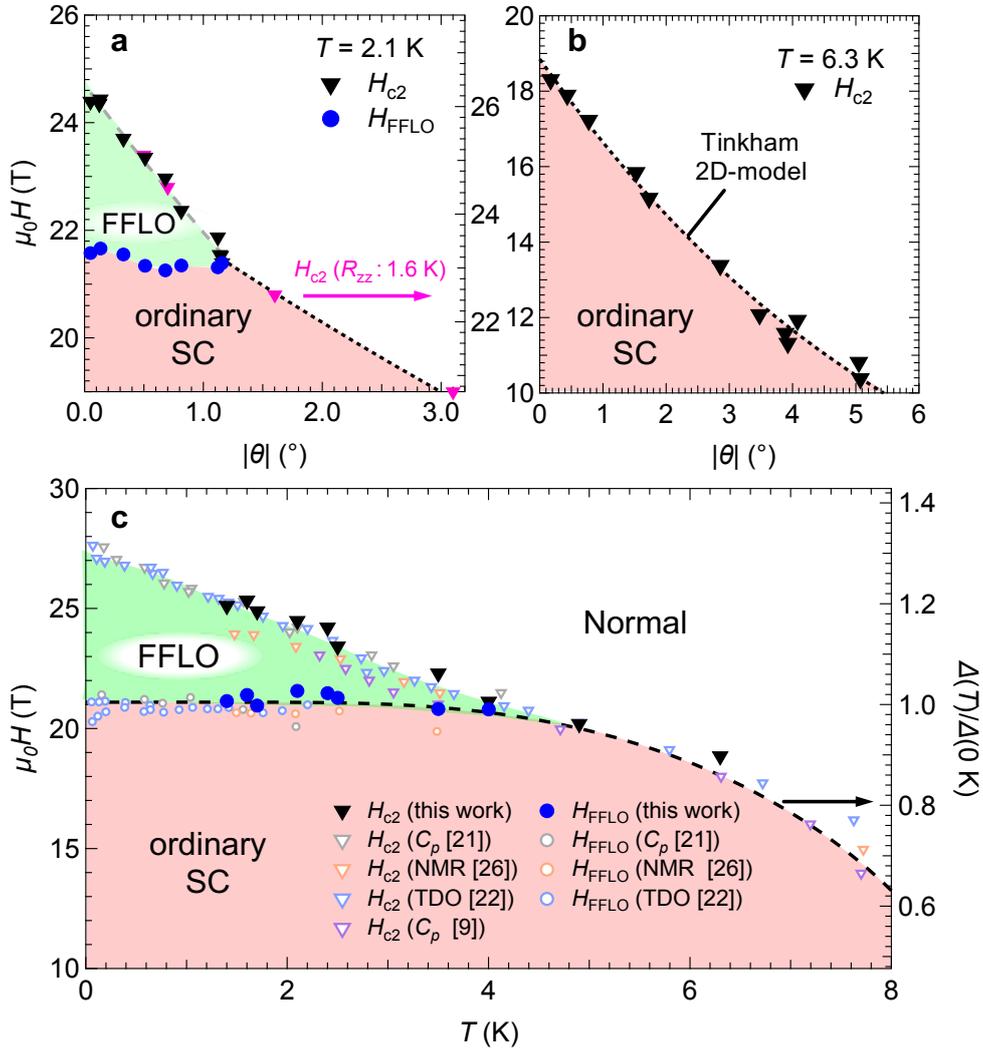

**Figure 4  Emergent anisotropy in the FFLO state.**

**a,b** $\Delta v/v$ at 1.6 K for $\boldsymbol{u}\|b$ (blue) and $\boldsymbol{u}\|c$ (red) in magnetic fields (**a**) parallel ($\theta$=0°) and (**b**) perpendicular ($\theta$=90°) to the 2D conducting plane. Direction dependence of the longitudinal sound wave propagation is observed only in the FFLO state (green area), whereas the direction dependence of $\Delta v/v$ in the BCS state and the normal state is not significant. The inset schematics show the electronic states (red: superconductivity, blue: normal state), directions of applied fields (light green arrows), and sound polarization vectors (striped arrows).

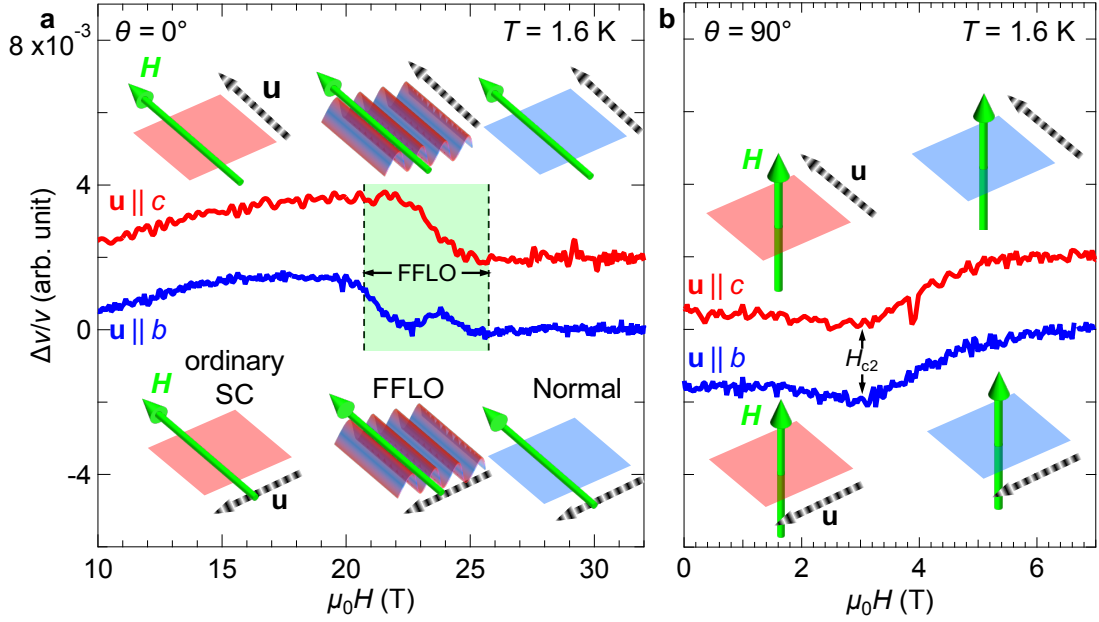

Shusaku Imajo[*], Toshihiro Nomura, Yoshimitsu Kohama, and Koichi Kindo

*Institute for Solid State Physics, University of Tokyo, Kashiwa 277-8581, Japan*

### 1. Details of our multidirectional ultrasound measurements

In this study, we implemented multidirectional ultrasound measurements using four LiNbO$_3$ piezoelectric transducers. Supplementary Figure 1a shows a photo of the sample setup for the measurements. The sample shape is indicated by the red dotted line, and the transducers are highlighted in blue. The gold wires in the picture are the electrical wiring to the transducers to apply and detect sound waves. Rotating magnetic fields are applied in the $a^*$-$c$ plane. As shown in Supplementary Fig. 1b, the crystal used in this study was sliced and formed into a square shape to attach the transducers with epoxy resin because typical crystals of κ-(BEDT-TTF)$_2$Cu(NCS)$_2$ were obtained in a hexagonal shape. Longitudinal sound waves were generated from one of the transducers and detected from another. The sound velocity $v$ is approximately 2.4 km/s at 10 K and has no significant in-plane anisotropy. Supplementary Figure 1c shows the temperature dependence of $\Delta v/v$ for the superconducting state (0 T) and the normal state (5 T, applied perpendicular to the conducting plane) of κ-(BEDT-TTF)$_2$Cu(NCS)$_2$. The data are well consistent with the reported data (1). Notably, measuring the temperature dependence with the above pulsed-field ultrasound measurement setup was difficult due to the temperature control, and therefore, only the temperature dependence was measured using another probe and another crystal in a static field.

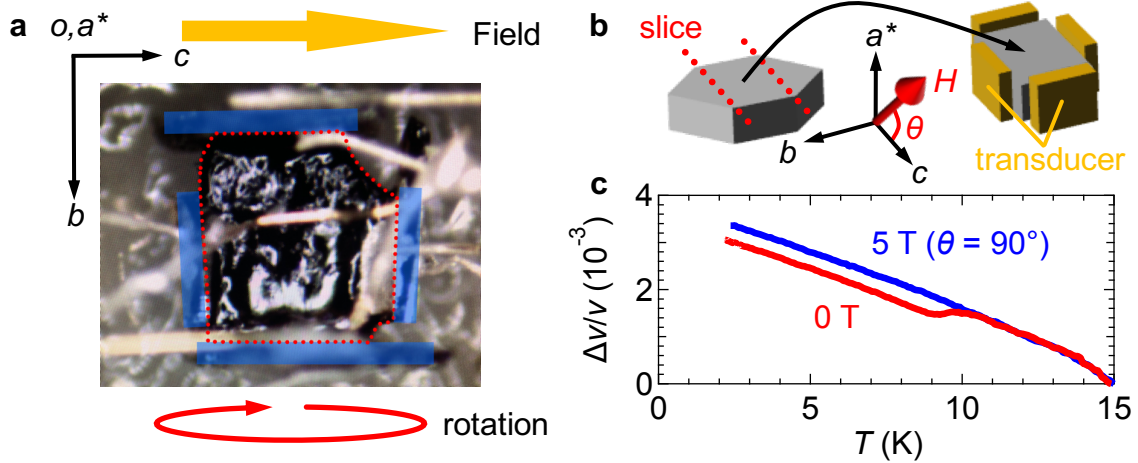

**Supplementary Figure 1: a** Photo of the sample setup with piezoelectric transducers (blue). The red dotted curve emphasizes the outline of the sample. Rotating magnetic fields are applied in the $a^*$-$c$ plane. **b** Schematic illustration of the treatment for attaching the transducers to the crystal. **c** Temperature dependence of $\Delta v/v$ at 0 T and 5 T when $\boldsymbol{u}\|b$ and $\boldsymbol{H}\|a^*$.

## 2. Pulsed magnetic field

  All the data in the main text are obtained by using pulsed magnetic fields, which were generated by an in-house pulse magnet. In Supplementary Fig. 2, field profiles of typical pulsed magnetic fields used in this study are shown. The duration of the pulsed fields was about 38 ms. Using a 0.9 MJ capacitor bank (18 mF), the 33.8 T and 51.9 T pulsed fields were generated with charging voltages of 5 kV and 8 kV, respectively.

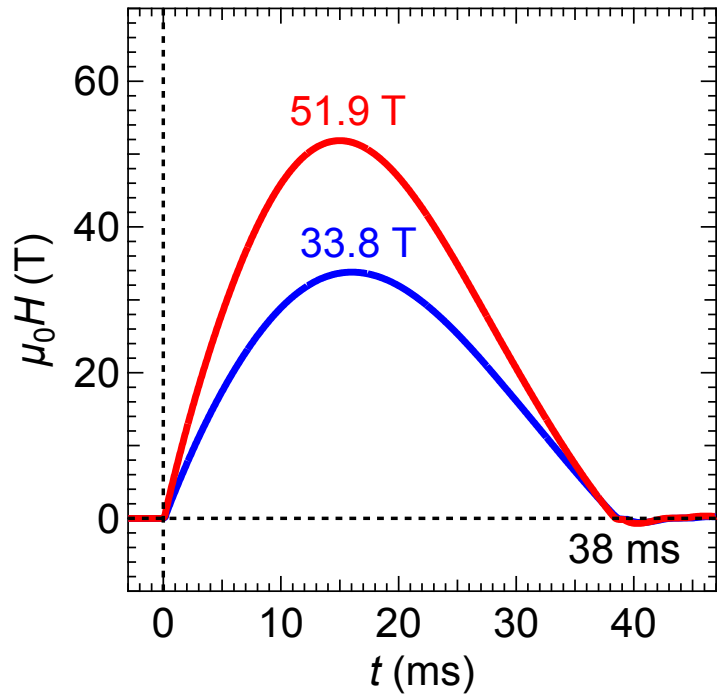

**Supplementary Figure 2:**  Field profiles of 33.8 T and 51.9 T pulsed-field shots with the duration of ~38 ms.

### 3. Acoustic de Haas oscillations

Supplementary Figure 3a shows the oscillatory part of $\Delta\alpha$, $\Delta\alpha_{osc}$, originating from the acoustic de Haas oscillation at 1.6 K. The black curve is a fit to the Lifshitz-Kosevich formula with the frequencies of the $\alpha$ orbit $F_\alpha$=610 T and the $\beta-\alpha$ orbit $F_{\beta-\alpha}$=3300 T. The Fourier spectrum of $\Delta\alpha_{osc}$ is shown in Supplementary Fig. 3b. This spectrum clearly indicates the absence of the $\beta$ orbit component with $F_\beta$=3900 T and the presence of the $\beta-\alpha$ orbit, known as the forbidden orbit, component in the acoustic de Haas signal. The absence of the $\beta$ orbit component is reasonable since the measurement temperature is relatively high ($T$>1.6 K) to detect the heavy-mass $\beta$ orbit ($m_\beta$~6$m_e$). Supplementary Figure 3c shows the mass plot in ln($A/T$) vs. $T$ format, where $A$ represents the Fourier amplitude estimated in the range from 30 T to 50 T. Both orbits show similar temperature dependence, and an effective mass of $m^*$~3.0$m_e$ is obtained for both. For the $\alpha$ orbit, the result shows good agreement with the reported data, $m_\alpha$~3.2$m_e$ (2,3). However, the value of $m_{\beta-\alpha}$ strongly depends on the measurement technique (2,3). The origin and features of the $\beta-\alpha$ orbit are still open questions.

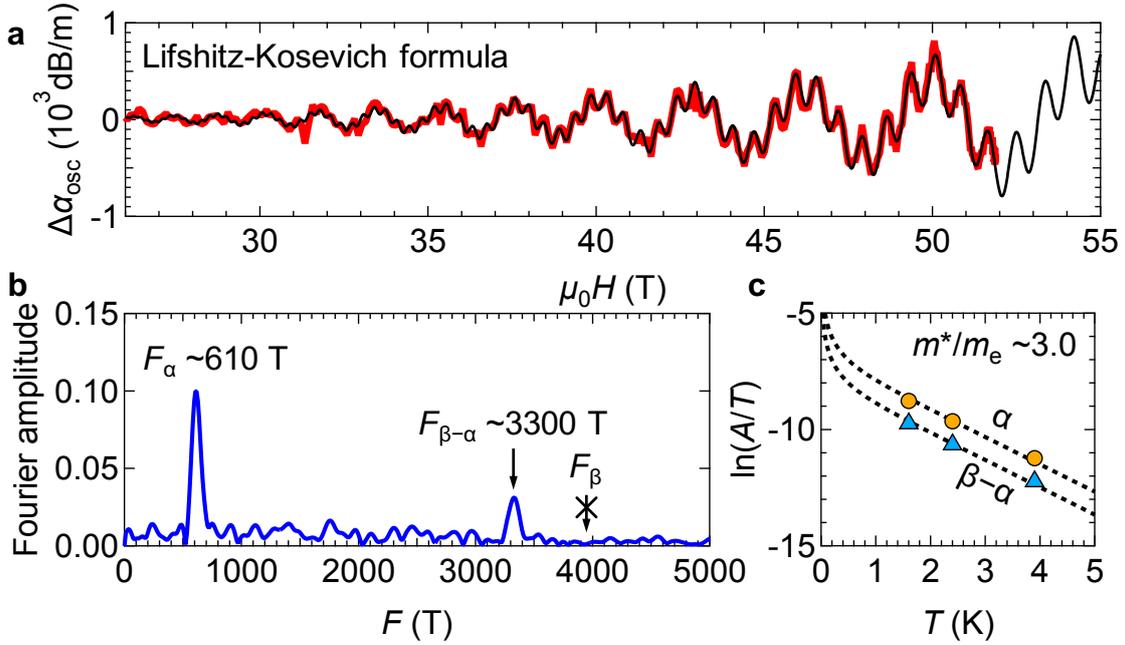

**Supplementary Figure 3: a** Acoustic de Haas oscillation observed in $\Delta\alpha$ at 1.6K. The black curve represents a fit to the two-component Lifshitz-Kosevich formula of the $\alpha$ orbit and the $\beta-\alpha$ orbit. **b** Fourier spectrum of the acoustic de Haas shown in **a**. The peaks are observed at the frequencies $F_\alpha$=610 T and $F_{\beta-\alpha}$=3300 T. The reported frequency for the $\beta$ orbit $F_{\beta-\alpha}$=3900 T (2) was not detected in the present measurement due to the relatively high temperature. **c** Mass plot of the detected orbits. Both are well reproduced with effective mass $m^*$~3.0$m_e$, which is consistent with the reported data $m_\alpha$~3.2$m_e$ (2,3).

## 5. Hysteresis at $H_{FFLO}$

The order parameter of the FFLO state includes the term of the **q** vector, cos(**qr**), which discontinuously jumps from zero to a finite value at $H_{FFLO}$. Therefore, the phase transition between the ordinary superconducting state and the FFLO state is a first-order one. Indeed, earlier studies demonstrate (4,5) that the transition at $H_{FFLO}$ in $\kappa$-(BEDT-TTF)$_2$Cu(NCS)$_2$ is a first-order transition showing a hysteresis. The hysteresis is clear only at very low temperatures, and becomes less clear above 1.5 K (4,5). Supplementary Figure 4 shows the ultrasound data at $\theta$=0° in up- and down-sweep fields at 1.6 K, which is the lowest temperature in our measurements. In these datasets, no clear hysteresis is observed, which is consistent with the previously reported data above 1.5 K.

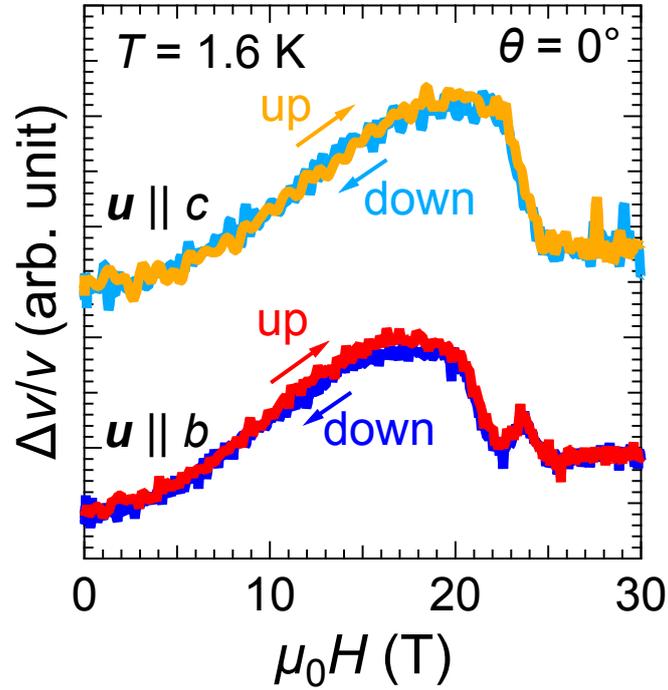

**Fig. S4** Field dependence of $\Delta v/v$ ($\theta$=0°) at the lowest temperature, 1.6 K, in up- (red and orange) and down-sweeps (blue and light blue).

## 5. Magnetic field dependence of $\Delta v/v$ in parallel fields at various temperatures

The data points $H_{c2}(\theta=0°)$ shown in Fig. 3 are obtained from the field dependence of $\Delta v/v$ at various temperatures shown in Supplementary Fig. 5. Above 4 K, an additional anomaly is observed at ~5 T. Since elastic properties are strongly influenced by vortices, this anomaly should be attributable to transitions of the vortices. The Josephson vortex lattices in 2D layered superconductors exhibit several transitions related to their depinning and melting, as reported (6). The drastic enhancement of $\Delta v/v$ below 5 T is most likely due to pinning of the flux lines since this anomaly disappears at low temperatures. This is because the interlayer coherence length in 2D superconductors becomes smaller than the interlayer spacing at lower temperatures, and the pinning effect on the Josephson vortices in the insulating layers is strongly suppressed due to interlayer decoupling. For depinned Josephson vortices, the flux-line lattice flows and can be easily deformed in the insulating plane; and therefore, the in-plane anisotropy should be reduced. This suppression of the in-plane anisotropy should correspond to the small sound wave direction dependence below $H_{\text{FFLO}}$, as shown in Fig. 4a.

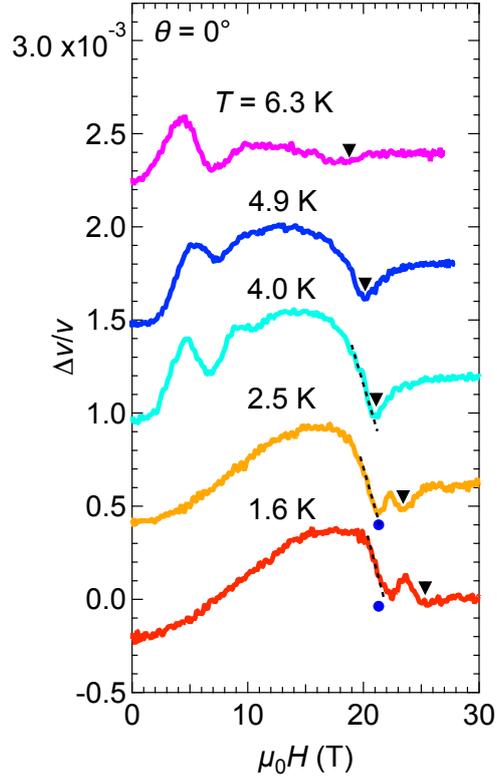

**Supplementary Figure 5:** In-plane magnetic field dependence of $\Delta v/v$ at various temperatures. The dashed lines are guides to make the transition at $H_{\text{FFLO}}$ (blue circle) clearer.

## 6. Out-of-plane electrical resistance

During the ultrasound measurements, we additionally measured the out-of-plane electrical resistance of another crystal at the same time. The field directions were the same as those shown in Supplementary Fig. 1 and electric currents were applied parallel to the $a^*$-axis. Supplementary Figure 6 shows the magnetic field dependence of the resistance at 1.6 K. Since these were just supplementary measurements obtained while performing the main ultrasound measurements, there are no data at $|\theta|=0°$. Nevertheless, the data at $\theta = -0.7°$ and $-0.5°$ show the resistivity data of the FFLO state appearing above $H_{FFLO}=21$ T. $H_{c2}$ is determined by the flection points shown by the pink triangles. As indicated by the arrows, these data show some kinks in the FFLO state. Since the finite resistance originates from the vortex dynamics, these kinks should be related to vortex pinning. Similar kink structures of the out-of-plane resistance are reported in other FFLO candidates (6,7) and are regarded as the commensurability effect of pinning on the FFLO spatial modulation (6-8). This effect occurs at certain fields where the FFLO wavelength $2\pi/q$ is commensurate with the Josephson vortex lattice constant $d_{JV}$, which leads to smaller resistances due to the relatively strong pinning at the nodes. This effect also indicates that the $\mathbf{q}$ vector is parallel to the $b$-axis because the spatial modulation needs to trap the flux lines, which is consistent with the results obtained by our ultrasound measurements.

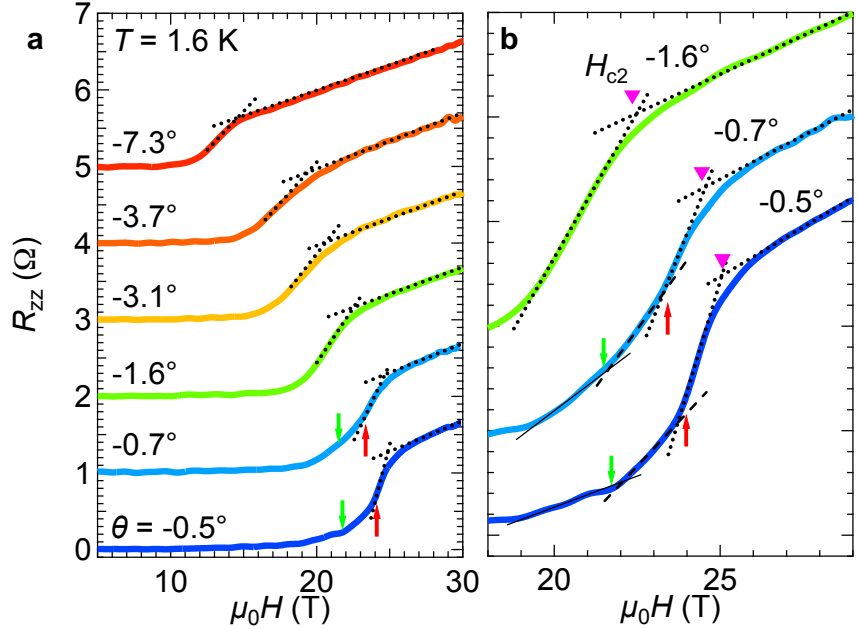

**Supplementary Figure S6: a,b** Out-of-plane magnetoresistance at 1.6 K at various angles. An enlarged plot of the datasets for $\theta=-1.6°$, $-0.7°$, and $-0.5°$ is shown in **b**. The pink triangle shows $H_{c2}$ at each angle. The arrows point to kinks in the FFLO state.

## Supplementary References